# In-plane and Out-of-plane Plasma Resonances in Optimally Doped La$_{1.84}$Sr$_{0.16}$CuO$_4$

Y. H. Kim[1*], P. H. Hor[2**], X. L. Dong[3], F. Zhou[3] and Z. X. Zhao[3]

[1]Department of Physics, University of Cincinnati, Cincinnati, Ohio 45221-0011, U. S. A.
[2]Texas Center for Superconductivity and Department of Physics
University of Houston, Houston, Texas 77204-5002, U. S. A.
[3]National Laboratory for Superconductivity and Beijing National Laboratory for Condensed Matter Physics, Institute of Physics, Chinese Academy of Sciences, P.O. Box 603, Beijing 100190, China

We addressed the inconsistency between the electron mass anisotropy ratios determined by the far-infrared experiments and *DC* conductivity measurements. By eliminating possible sources of error and increasing the sensitivity and resolution in the far-infrared reflectivity measurement on the single crystalline and on the polycrystalline La$_{1.84}$Sr$_{0.16}$CuO$_4$, we have unambiguously identified that the source of the mass anisotropy problem is in the estimation of the free electron density involved in the charge transport and superconductivity. In this study we found that only 2.8 % of the total doping-induced charge density is itinerant at optimal doping. Our result not only resolves the mass anisotropy puzzle but also points to a novel electronic structure formed by the rest of the electrons that sets the stage for the high temperature superconductivity.

*kimy@ucmail.uc.edu
**pei@Central.UH.EDU

Our understanding of the high temperature superconductivity (HTS) in copper oxide materials (cuprates) has always been refined by continuous improvements on the sample qualities and the experimental techniques. For instance, while numerous angle-resolved photoelectron spectroscopy (ARPES) data have been published, only after the sample was probed using the low energy photon with an energy resolution of ~ 3 meV, it was finally shown that the doping dependence of the Fermi velocity obtained by ARPES indeed agrees with that of the thermal conductivity measurement [1]. Therefore, in order to bring about a coherent picture of the HTS, it is of fundamental importance to resolve the inconsistency in essential physical quantities that should be independent of the experimental probes.

There is one perplexing disagreement between the *DC* and *AC* transport measurements of underdoped cuprates that has long been overlooked. Despite the general consensus that the superconducting cuprate is metallic in the $CuO_2$ plane (ab-plane) but insulating in the direction perpendicular to the ab-plane (c-axis), which leads to a cylindrical Femi surface, recent quantum oscillations measurements of $Y_1Ba_2Cu_3O_{6.5}$ [2] and $Y_1Ba_2Cu_4O_8$ [3], which directly probe the free electrons (or holes) in the $CuO_2$ plane, found on the one hand that the area of the two-dimensional (2D) Fermi surface is only 2 and 2.4 % of the total area of the Brillouin zone respectively, corresponding roughly to ~ 3 % of total doping-induced electrons when compared with the large Fermi surface observed in the over-doped regime [2]. On the other hand, the reported optical plasma frequency falls in the range of ~ 1.0 eV for all cuprates and is peculiarly insensitive to the doping levels ranging from underdoped to overdoped regimes [4].

The ab-plane optical reflectivity data published by many groups shows the metallic reflectivity with a plasma edge at ~ 0.8 – 1.0 eV for LSCO [4 – 6], ~ 1.0 eV – 1.2 eV for YBCO [7, 8], ~ 1.0 eV for $Bi_2Sr_2CaCu_2O_8$ (Bi-2212) [9], and ~ 1.2 eV $Tl_2Ba_2Ca_2Cu_3O_8$ [10, 11].

However the c-axis plasma edge that emerges out of an insulator-like flat reflectivity only when $T < T_c$ was found at $\sim 50$ cm$^{-1}$ (6 meV) – 85 cm$^{-1}$ (10 meV) for optimally doped LSCO [12 – 14], $\sim 65$ cm$^{-1}$ (8 meV) for YBCO$_{6.7}$ [15], $\sim 250$ cm$^{-1}$ (30 meV) for YBCO$_{6.95}$ [16], $\sim 220$ cm$^{-1}$ (27 meV) for Y$_1$Ba$_2$Cu$_4$O$_8$ [17], $\sim 40$ cm$^{-1}$ (5 meV) for Bi$_{1.6}$Pb$_{0.6}$Sr$_{1.8}$CaCu$_2$O$_y$ ($T_c$ = 65 K) [18], $\sim 10$ cm$^{-1}$ (1.2 meV) for Bi-2212 ($T_c \sim 90$ K) [19], and $\sim 40$ cm$^{-1}$ (5 meV) for Tl$_2$Ba$_2$Ca$_2$Cu$_3$O$_{6+x}$ ($T_c$ = 81 K) [11]. Since superconductivity is intrinsically a three-dimensional (3D) phenomenon, the screened ab-plane plasma frequency $\tilde{\omega}_{ab}^2 = \omega_{ab}^2/\varepsilon_{ab} = 4\pi n_{ab}e^2/\varepsilon_{ab}m_{ab}$ and that of the c-axis $\tilde{\omega}_c^2 = 4\pi n_c e^2/\varepsilon_c m_c$ found through the optical experiments must arise from the same free electrons (*i. e.*, $n_{ab} = n_c$ at $T = 0$ K). Therefore, the mass anisotropy of the free electrons ($m_c/m_{ab}$) may be calculated from $\tilde{\omega}_{ab}^2/\tilde{\omega}_c^2$ to yield $\sim 690000$ for Bi-2212, $\sim 10000$ for LSCO, and $\sim 1600$ for YBCO$_{6.95}$, with $\varepsilon_{ab} \approx \varepsilon_c$ which is estimated typically in the range of $\sim 12 - 40$. However, the *DC* transport measurements found that the mass anisotropy ranges from $\sim 10000$ for Bi-2212 to $\sim 200$ for LSCO, and to $\sim 10$ for Y$_1$Ba$_2$Cu$_3$O$_7$ [20].

From the canonical charge transport point of view, the mass anisotropy found by the optical measurements should be the same as that determined by the *DC* transport measurements in the long wavelength limit. Thus, the disagreement between the optical and *DC* transport measurements poses a serious problem at the foundational level. This discrepancy must have come from the initial assumptions of either $n_{ab} = n_c$ or $\varepsilon_{ab} \approx \varepsilon_c$ or both when the plasma frequency ratio is considered. Since we do not expect much anisotropy between $\varepsilon_{ab}$ and $\varepsilon_c$ as the LDA calculation suggests [21], therefore, we are left with $n_{ab}/n_c \sim 50$ for LSCO at x = 0.16 for example. This suggests that the reported free electron plasma edge in the ab-plane optical reflectivity of LSCO at $\sim 1$ eV must involve $\sim 50$ times more free electrons than those that are actually responsible for the *DC* conductivity.

In this work we measured the far-IR ab-plane reflectivity of the optimally doped ($T_c = 37$ K) single crystalline LSCO (S-LSCO) of area ~ 5 x 5 mm$^2$ and the reflectivity of optimally doped ($T_c = 38$ K) polycrystalline LSCO (P-LSCO) of area ~ 1 cm$^2$ for comparison. An unpolarized far-IR beam was used because no appreciable difference between the π–polarized (parallel to the plane of incidence) and the σ-polarized (perpendicular to the plane of incidence) reflectivity at an 8° angle of incidence was observed [22]. A Bruker 113v spectrometer was used with a composite Si bolometer operating at 4.2 K for frequencies between ~ 20 cm$^{-1}$ and 400 cm$^{-1}$. In order to cover the frequencies below 30 cm$^{-1}$, a composite Cu-doped Si bolometer with a 1 cm$^2$ active area operating at 2 K was used in conjunction with a parabolic light cone with 7 mm diameter exit aperture and a 75 μm Mylar beam splitter.

As the reference for $\omega < 400$ cm$^{-1}$ a gold mirror was used and an aluminum mirror for the range between 400 cm$^{-1} < \omega < 4000$ cm$^{-1}$. The sample temperature was directly monitored from the backside of the sample. The spectral resolution was 1 cm$^{-1}$ for $\omega < 120$ cm$^{-1}$, 2 cm$^{-1}$ for 120 cm$^{-1} < \omega < 400$ cm$^{-1}$, and 4 cm$^{-1}$ for $\omega > 400$ cm$^{-1}$. Our laser sample positioning set-up minimized the error in the reflectivity to less than ± 0.5 % error in establishing the precise optical alignment as the reference mirror and the sample interchanged. The real part of the conductivity ($\sigma_1$) and the real part of the dielectric function ($\varepsilon_1$) were calculated using a Kramers-Kronig transformation of the reflectivity data. For the high frequency extrapolation we have used the optical data reported [4] whereas the Hagen-Ruben's rule was used for the low frequency extrapolation. It turns out that the spectral features presented in this paper are insensitive to the extrapolations outside the spectral range of interest.

The reflectivity data of S-LSCO and P-LSCO taken as a function of temperature are shown in Fig. 1. Except for smoothing of the S-LSCO data for $\omega < 25$ cm$^{-1}$ and the P-LSCO data for $\omega$

< 18 cm$^{-1}$, raw reflectivity data is displayed and used for data analysis. In the reflectivity plot of S-LSCO two things can be noticed: (1) While the overall far-IR reflectivity is high above 90 %, it never reaches 100% even in the superconducting state except for frequencies below 25 cm$^{-1}$ within the experimental uncertainty and (2) the reflectivity shows a broad local minimum at around ~ 90 cm$^{-1}$ indicated with a red arrow. In the P-LSCO reflectivity it is clear that overall reflectivity for ω < 200 cm$^{-1}$ is qualitatively the same as that of S-LSCO, which proves that far-IR study of the polycrystalline cuprate indeed offers an advantage over that of the single crystalline sample due to its lower overall reflectivity and larger sample size with the well-documented anisotropic spectral information. Furthermore, as a byproduct, the reflectivity of P-LSCO shows the development of the c-axis plasma (so-called Josephson plasma) dip at ~ 89 cm$^{-1}$ (blue arrow) for $T \leq T_c$. However, in contrast to the previous report [23], the position of the dip does not appear to move progressively from zero to 89 cm$^{-1}$ with decreasing $T$. Instead, its depth grows deeper while maintaining the same frequency as $T \rightarrow 0$. In addition, there exists a small dip at ~ 92 cm$^{-1}$ in the reflectivity of P-LSCO even at room temperature as also indicated with a red arrow.

Fig. 2 displays $\sigma_1$ and $\varepsilon_1$ of S-LSCO and P-LSCO side by side. The c-axis infrared-active $A_{2u}(3)$, $A_{2u}(2)$, and $A_{2u}(1)$ modes [24] survive the screening, and appear respectively at $C_1$ ~ 220 cm$^{-1}$, $C_2$ ~ 350 cm$^{-1}$, and $C_3$ ~ 500 cm$^{-1}$ in the P-LSCO plot. On the contrary, all three $E_u$ modes of the ab-plane [24] are screened out upon doping and the new ab-plane doping-induced modes indicated as $X_1$ and $X_2$ appear instead as a result of the symmetry breaking of the Raman-active $E_{1g}(2)$ and $E_{1g}(1)$ ab-plane modes respectively [25, 26]. However the additional ab-plane doping-induced modes $\omega_{G1}$ ~ 23 cm$^{-1}$, $\omega_{G2}$ ~ 36 cm$^{-1}$, and $\omega_{G3}$ ~ 72 cm$^{-1}$ do not have their counterparts in the Raman spectra of the $CuO_2$ lattice as pointed out in Ref [26, 27]. Moreover, the presence of the intense $X_1$ peak in S-LSCO has been the source of confusion [6, 26] although there has been ample hint of its presence [28 – 30]. Notice that $X_1$ at ~ 110 cm$^{-1}$ appears only as a small gradual downward step at ~ 110 cm$^{-1}$ with decreasing frequency in the reflectivity of S-LSCO (see Fig. 1).

The $\varepsilon_1$ plot shows that the free electron contribution to $\varepsilon_1$, which is characterized by the negative $\varepsilon_1$ for $\omega < \tilde{\omega}_{ab}$, is confined to the frequency range below 100 cm$^{-1}$ for both S-LSCO and P-LSCO samples. The spectroscopy of the energy loss function, $\text{Im}(-1/\varepsilon) = \varepsilon_2/(\varepsilon_1^2 + \varepsilon_2^2)$ which is sensitive to the zeroes of $\varepsilon_1(\omega)$ and the results are displayed in Fig. 3. In addition to the peaks related to $\omega_{G1}$, $\omega_{G2}$, and $\omega_{G3}$ modes, the fourth mode denoted as $\omega_{ab}^L$ at ~ 92 cm$^{-1}$, which is present in both S-LSCO and P-LSCO, must originate from $\tilde{\omega}_{ab}$ ~ 54 cm$^{-1}$ via $\omega_{ab}^L \approx \sqrt{\tilde{\omega}_{ab}^2 + \omega_{G3}^2}$ due to the presence of the resonance at $\omega_{G3}$ ~ 72 cm$^{-1}$ which has been discussed in detail in Ref. [31]. In addition, the loss function of P-LSCO (upper panel) shows the development of a peak between the energy loss peaks of $\omega_{G3}$ and $\omega_{ab}^L$, indicated as $\tilde{\omega}_c$ at ~ 85 cm$^{-1}$ that becomes slightly blue-shifted to ~ 89 cm$^{-1}$ as $T \rightarrow 0$.

Further insight into the free electron plasma frequency can be gained from the two fluid model $\varepsilon \equiv \varepsilon_1 + i\varepsilon_2 = \varepsilon_\infty - f_s \omega_p^2/\omega^2 - (1 - f_s)\omega_p^2/\omega(\omega - i\Gamma)$ where $\varepsilon_\infty$ is the dielectric constant responsible for the screening, $\omega_p$ is the unscreened plasma frequency, $f_s$ is the superfluid fraction, and $\Gamma$ is the scattering rate of the electrons in normal state. Thus in $T \rightarrow 0$ ($f_s \rightarrow 1$) limit, we have $\varepsilon_1 \approx \varepsilon_\infty - \omega_p^2/\omega^2$. Hence, by plotting $\varepsilon_1$ at $T = 14$ K (our lowest temperature) versus $\omega^{-2}$, the unscreened $\omega_p^2$ may be found from the slope and $\varepsilon_\infty$ from the y-intercept. As shown in Fig. 4 (a), we obtain $\varepsilon_\infty = \varepsilon_{ab}$ ~ 1600 and $\omega_{ab}$ ~ 2000 cm$^{-1}$ for S-LSCO. Thus $\tilde{\omega}_{ab} = \omega_{ab}/\sqrt{\varepsilon_{ab}}$ ~ 50 cm$^{-1}$ which is consistent with $\tilde{\omega}_{ab}$ ~ 54 cm$^{-1}$ found from the energy loss function. Since P-LSCO data shown in Fig. 4(b) contains ~ 2% ab-plane contribution [26], the c-axis portion becomes $\varepsilon_c \sim 69$ and $\omega_c \sim 735$ cm$^{-1}$ from the $\varepsilon_\infty \sim 100$ and $\omega_p \sim 760$ cm$^{-1}$ of the linear fit, which yield $\tilde{\omega}_c \sim 88$ cm$^{-1}$ that is also in agreement with $\tilde{\omega}_c$ at ~ 89 cm$^{-1}$ of Fig. 3. Thus, comparing $\omega_{ab} \sim 2000$ cm$^{-1}$ found in this work with the theoretical plasma frequency $\omega_{ab}^{th} \sim 12000$ cm$^{-1}$ of the electrons of density $n = 1.6 \times 10^{21}$ electrons/cm$^3$ at x = 0.16, the fraction of the actual free electrons is $n_{ab}/n \sim 0.028$.

Since the width and strength of the energy loss peak is governed by $\varepsilon_2$, we expect $\omega_{ab}^L$ to disappear in the superconducting state as $\varepsilon_2 = 4\pi\sigma_1/\omega = 0$. Indeed the $T$-dependence of $\omega_{ab}^L$ demonstrates the disappearance of the energy loss peak at $\omega_{ab}^L$ in S-LSCO as shown in Fig. 4(c). Notice that $\omega_{ab}^L$ in P-LSCO (see Fig. 3) also vanishes as $T \rightarrow 0$, confirming the same physical origin as the $\omega_{ab}^L$ in S-LSCO. However, while the $\omega_{ab}^L$ disappears for T < T$_c$, $\tilde{\omega}_c \sim 89$ cm$^{-1}$ persists in the superconducting state, which implies that the $\varepsilon_2$ in the c-axis is not zero even for T < T$_c$ suggesting that not all the free electrons are superconducting along the c-axis. This was pointed out in the previous c-axis far-IR study of LSCO [32] where the c-axis superfluid density fraction of x = 0.16 LSCO was found to be 87 % of the free electron density. This c-axis superfluid fraction rapidly decreases with decreasing x reaching down to only 22 % at x = 0.07 [32].

Furthermore it was found that the c-axis scattering rate $\Gamma_c$ increases linearly with x from only ~ 13 cm$^{-1}$ at x = 0.07 to ~ 170 cm$^{-1}$ at x = 0.16. The origin of the increasing $\Gamma_c$ with doping was attributed to the Coulomb scattering with the Sr-dopant ions between the CuO$_2$ planes [32]. Considering the fact that the scattering rate of Cu, $\Gamma_{Cu} \sim 160$ cm$^{-1}$ at 77 K, the observed $\Gamma_c \sim 170$ cm$^{-1}$ at x = 0.16 [12] is comparable to that of Cu. This suggests that there must exist a band-like charge transport channel along the c-axis that does not provide the pairing glue for the electrons and the c-axis electron transport is intrinsically metallic. This is in contrast to the common notion of the insulating behavior along the c-axis, which was conjectured based on the absence of the c-axis plasma edge in the normal state far-IR reflectivity and by the $T$-dependence of the c-axis resistivity. Since the fraction of the free electrons becomes less than 1 % in underdoped S-LSCO [22] that increases to ~ 2.8 % at optimal doping as found in this work, the c-axis screened plasma frequency $\tilde{\omega}_c$ is smaller than $\Gamma_c$, making the c-axis plasma oscillation appear over-damped. In fact the c-axis plasma edge was observed in the normal state c-axis reflectivity for x = 0.07 but not for x ≥ 0.08 even though the c-axis transport is metallic [32]. Therefore, the apparent hopping-like c-axis resistivity in underdoped LSCO may be attributed to the extrinsic effect arising from the stacking disorder introduced during the crystal growth along the c-axis. Indeed a metallic c-axis transport behavior was observed in 1 – 3 micron thick Bi-2122 samples [33].

The ab-plane scattering rate ($\Gamma_{ab}$) of the free electrons can be found by fitting of the low frequency tail of $\sigma_1$ of S-LSCO to $\sigma_{dc}$ of the same sample by using a Drude formula. As shown in Fig. 4(d), it was found that $\Gamma_{ab}$ ~ 12 cm$^{-1}$ at 300 K and 7 cm$^{-1}$ at 50 K. Hence, $\omega_{ab}$ may be calculated independently by $\omega_{ab} = \sqrt{60\sigma_{dc}\Gamma_{ab}}$ with $\sigma_{dc}$ = 1.06 x 10$^4$ $\Omega^{-1}$cm$^{-1}$ to find $\omega_{ab}$ ~ 2100 cm$^{-1}$ at T = 50 K, which is in good agreement with $\omega_{ab}$ ~ 2000 cm$^{-1}$ found from the two-fluid model. Thus from our far-IR observation, the intrinsic free electron mass anisotropy in x = 0.16 LSCO is $m_c/m_{ab} = (\omega_{ab}/\omega_c)^2$ ~ 10. Thus, the transport anisotropy measurement tends to overestimate the ratio due to the aforementioned extrinsic c-axis disorder effect.

In summary, we studied the charge dynamics of optimally doped LSCO in both the single crystalline and the polycrystalline morphologies. We found that the ab-plane normal state transport is nearly dissipationless and the c-axis scattering rate is comparable to that of copper, making it an anisotropic 3D metal with a mass anisotropy of the order of 10. Our observation suggests that not only 2.8 % of the doping-induced electrons are itinerant, which is consistent with the quantum oscillations measurements, but also the free electrons in the ab-plane suffer very little scattering ($\Gamma_{ab} \sim 7$ cm$^{-1}$ at T = 50 K) in an environment set up by the rest of the electrons that are responsible for the ab-plane charge-induced far-IR modes and the massive screening ($\varepsilon_{ab} \sim 1600$) of the free electrons. This observation, therefore, paints a characteristically new electronic picture that calls for a model which will lead to the right answer for the mechanism of HTS.

**Acknowledgement**

P.H.H. is supported by the State of Texas through the Texas Center for Superconductivity at University of Houston. X.L.D., F.Z. and Z.X.Z. are supported by Project No. 10874211 supported by NSFC and the National Basic Research Program of China (2011CB921703).

**Figure Captions:**

Fig. 1. Far-infrared reflectivity spectra of optimally doped S-LSCO and P-LSCO at various temperatures. Top to bottom: T = 14 K (blue), 20 K (blue), 25 K (blue), and 30 K (blue); T = 32 K (cyan), 35 K (cyan), 40 K (cyan), and 45 K (cyan); T = 50 K (green), 70 K (green), 100 K (green), 120 K (green), 150 K (green), 180 K (green), and 200 K (green); T = 250 K (orange) and 300 K (red).

Fig. 2. The corresponding frequency-dependent conductivity and dielectric function calculated from the reflectivity in Fig. 1. The same color scheme as in Fig.1 is used. See the text for details.

Fig. 3. The energy loss function of S-LSCO and P-LSCO at various temperatures (the same color scheme as in Fig. 1). See the text for details.

Fig. 4. The real part of the dielectric function as a function of $\omega^{-2}$ of S-LSCO at 14 K (Panel (a)) and P-LSCO at 14 K ( Panel (b)). The straight lines are linear fit. A close-up plot of the $\omega_{ab}^{L}$ peak in the energy loss function at T = 50 K (green), 40 K (thick cyan), 20 K (thick blue), and 14 K (blue) is shown in Panel (c) and the Drude fit of the low frequency tail of the real part conductivity to the corresponding dc conductivity at T = 300 K (red) and 50 K (cyan) is shown in Panel (d).

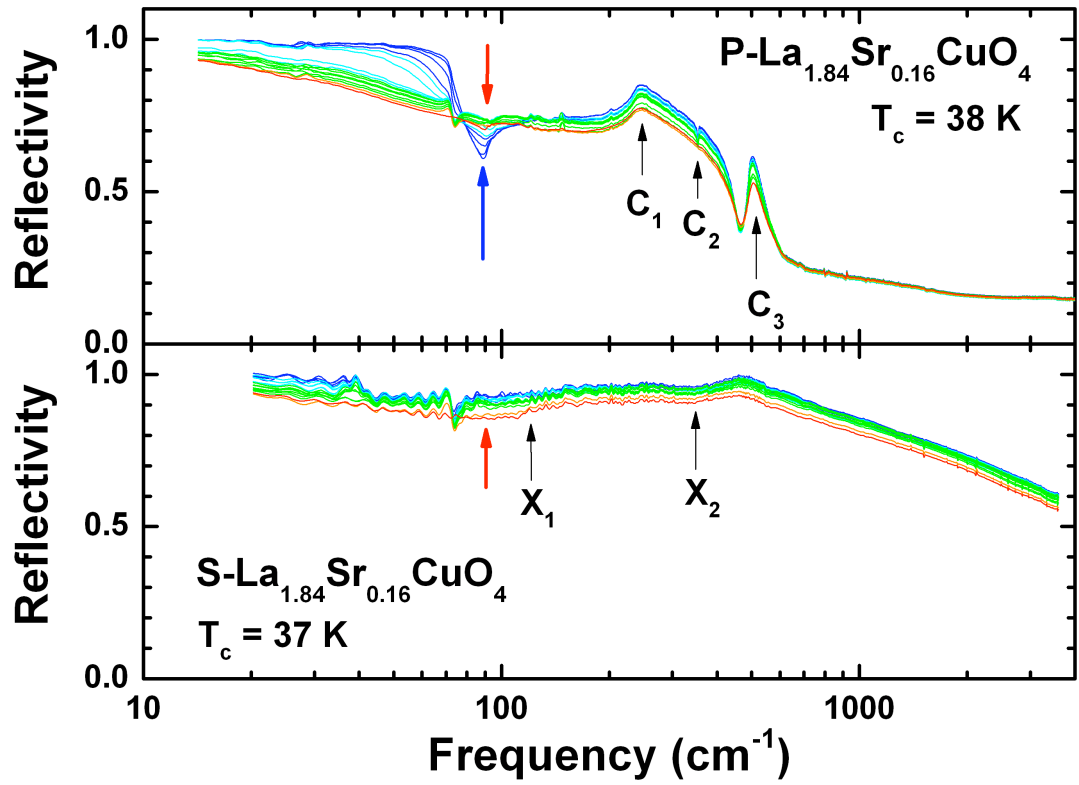

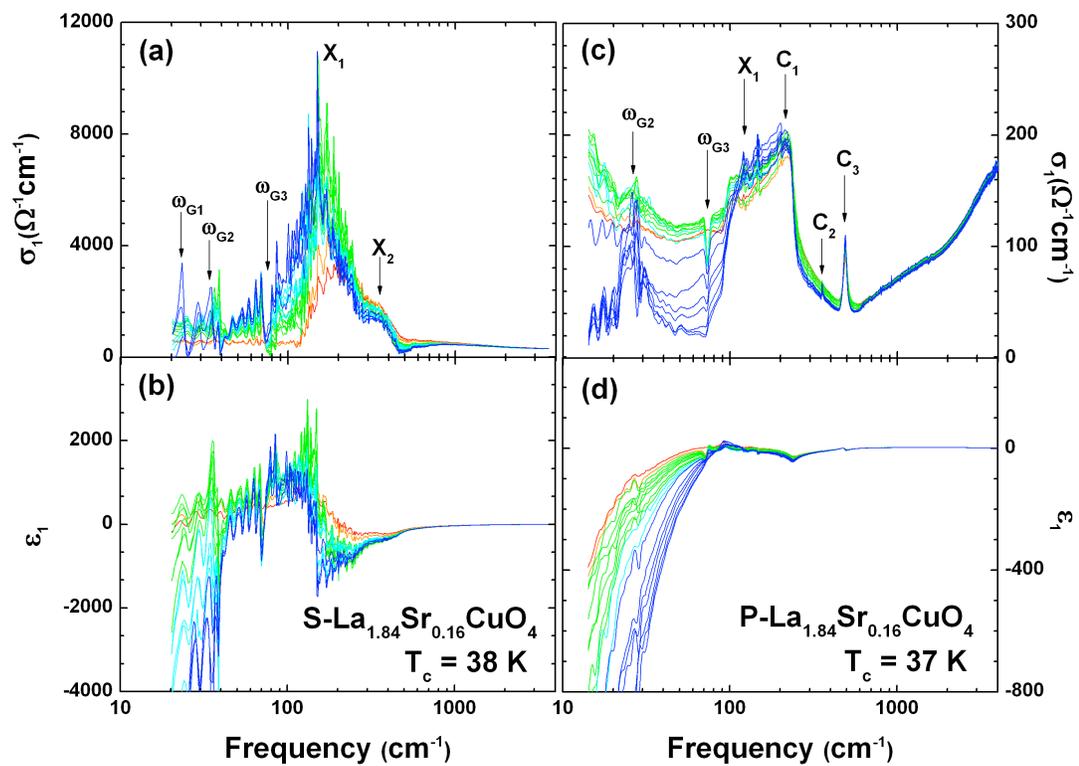

Figure 2

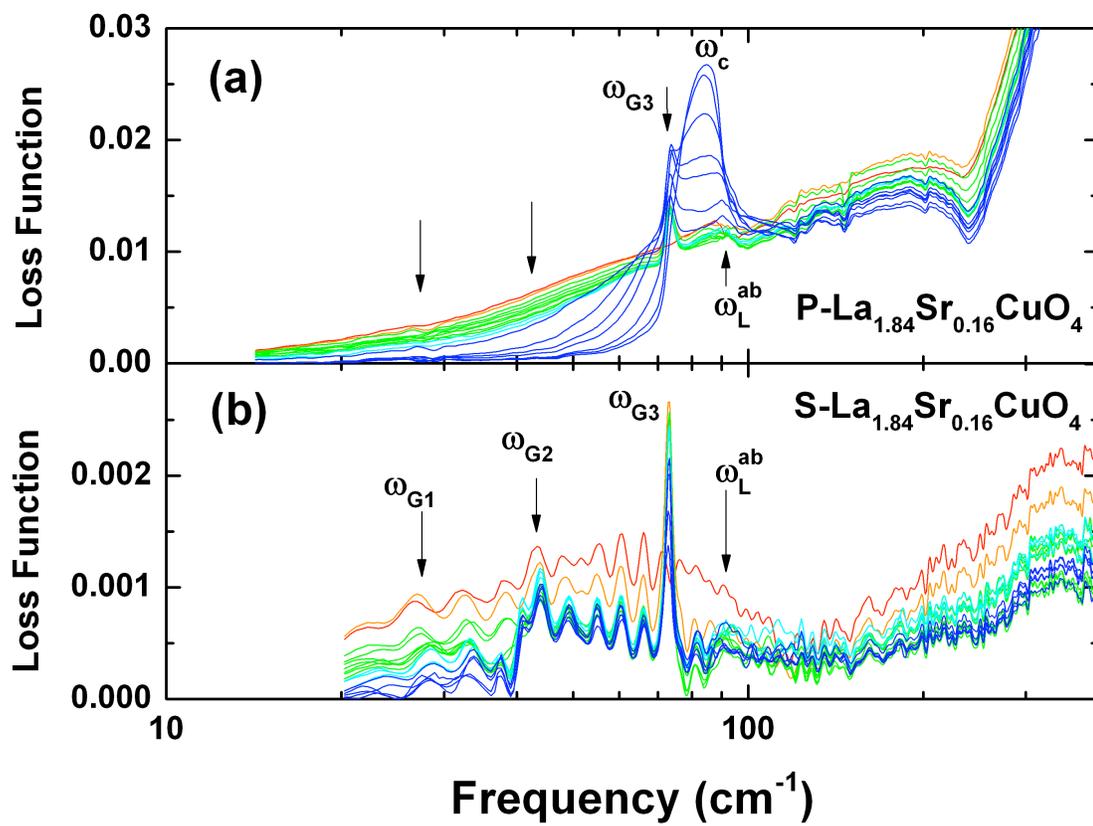

Figure 3

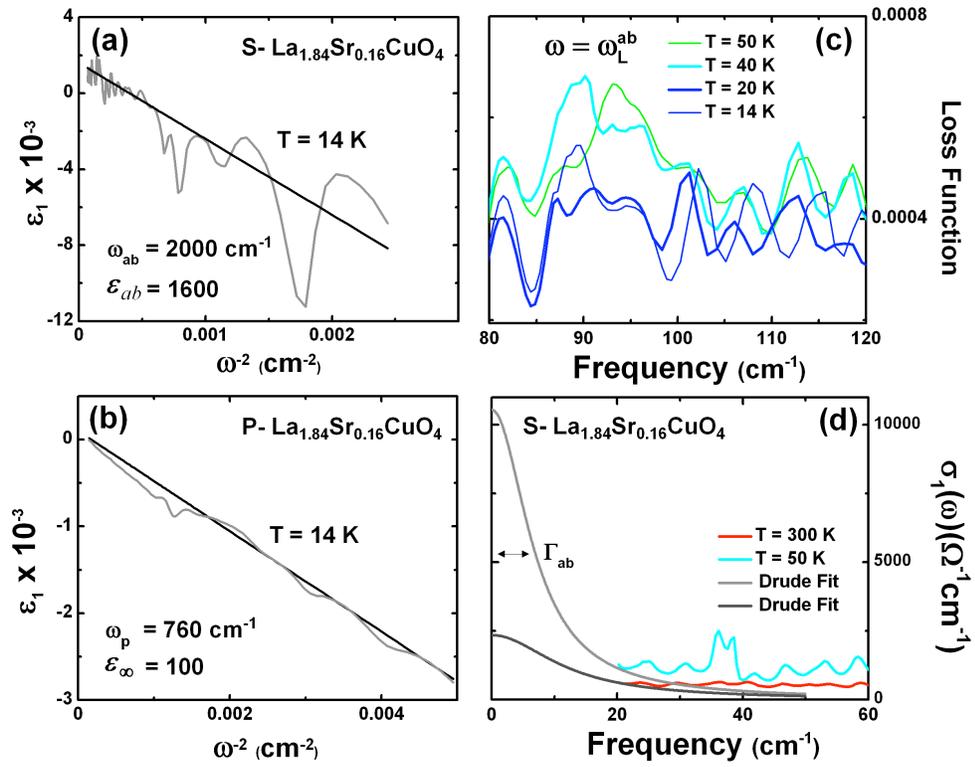

Figure 4